\documentclass[10pt,journal]{IEEEtran}
\usepackage[letterpaper, top=0.6in, bottom=0.7in, left=0.45in, right=0.45in, columnsep=0.15in]{geometry}
\IEEEoverridecommandlockouts

\usepackage[utf8]{inputenc}
\usepackage[T1]{fontenc}
\usepackage{url}
\usepackage{ifthen}
\usepackage{cite}
\usepackage{amsmath} 
\usepackage{physics}
\usepackage{array}
\usepackage{nicematrix}                          
\usepackage{amsthm}
\usepackage{amsfonts}
\usepackage{mathrsfs}
\usepackage{multirow}
\usepackage{citesort}
\usepackage{adjustbox}
\usepackage{tikz}
\usetikzlibrary{arrows.meta}
\usepackage{pgfplots}
\pgfplotsset{compat=1.18}
\usepackage{bm}
\usepackage{algorithm}
\usepackage{algpseudocodex}
\usetikzlibrary{patterns}
\usetikzlibrary{matrix,decorations.pathreplacing}
\usetikzlibrary{decorations.pathreplacing,calligraphy}
\usepackage{graphicx}
\usepackage{epstopdf}
\usepackage{amssymb}
\usepackage{enumitem}

\usepackage{diagbox}
\usepackage{makecell}
\usepackage{siunitx}
\usepackage{bigstrut}
\usepackage{tikz}
\usepackage{multirow}
\usepackage{booktabs}
\usepackage{stmaryrd}
\usepackage[caption=false]{subfig}
\usepackage{amsmath,amssymb,mathtools}

\usepackage{xhfill}
\allowdisplaybreaks[4]
\newcommand{\xfilll}[2][1ex]{%
	\dimen0=#2\advance\dimen0 by #1%
	\leaders\hrule height \dimen0 depth -#1\hfill%
}

\newtheoremstyle{thry}
{3pt}
{3pt}
{}
{1em}
{}
{:}
{.5em}
{}
\theoremstyle{thry}
\newtheorem{definition}{\textbf{Definition}}
\newtheorem{proposition}{\emph{\textbf{Proposition}}}

\newtheorem{example}{\emph{Example}}
\IfFileExists{Figures/concatenate.tex}
{
  
}
{
  
}

\ifCLASSOPTIONonecolumn
\newcommand{\figwidth}{0.65\textwidth}

\fi

\ifCLASSOPTIONtwocolumn
\newcommand{\figwidth}{0.42\textwidth}

\fi

\begin{document}
\title{Guessing Decoding of Short Blocklength Codes}
 \author{
\IEEEauthorblockN{Qianfan Wang,~\IEEEmembership{Member,~IEEE}, Jifan Liang, Peihong Yuan,~\IEEEmembership{Member,~IEEE}, Ken R. Duffy,~\IEEEmembership{Senior Member,~IEEE}, Muriel M\'edard,~\IEEEmembership{Fellow, IEEE}, and Xiao Ma,~\IEEEmembership{Member,~IEEE}
\thanks{Qianfan Wang is with the Department of
Computer Science, City University of Hong Kong, Hong Kong SAR, China~(e-mail: wangqf26@mail.sysu.edu.cn).}
\thanks{Peihong Yuan is with the Institute of Space Internet, Fudan University,
Shanghai 200433, China (e-mail: phyuan@fudan.edu.cn).}
\thanks{Muriel M\'edard is with the Network Coding and Reliable
Communications Group, Massachusetts Institute of Technology, Cambridge,
MA 02139 USA (e-mail: medard@mit.edu).}
\thanks{Ken R. Duffy is with the Engineering Probability Information and Communications Laboratory, Northeastern University, Boston, MA 02115 USA (e-mail: k.duffy@northeastern.edu)}
\thanks{Jifan Liang and Xiao Ma are with the School of Computer Science and Engineering, Sun Yat-sen University, Guangzhou 510006, China (e-mail: liangjf56@mail2.sysu.edu.cn, maxiao@mail.sysu.edu.cn). (Corresponding author: Xiao Ma.)}
}}

%
%

\maketitle
\pagestyle{empty}
\thispagestyle{empty}

\begin{abstract}
Future beyond-5G and 6G systems demand ultra-reliable, low-latency communication with short blocklengths, motivating the development of universal decoding algorithms.  Guessing decoding, which infers the noise or codeword candidate in order of decreasing~(exact or approximate) likelihood, offers a universal framework applicable to short codes.  In this paper, we present a unified treatment of two prominent recent families of guessing decoding: guessing random additive noise decoding~(GRAND) and  guessing codeword decoding~(GCD).  For each, we (i) present algorithmic implementations and ordering strategies; (ii) prove maximum-likelihood~(ML) optimality under appropriate stopping criteria; (iii) derive saddle-point approximations for the average number of queries; and (iv) validate theoretical predictions with simulations.
We further analyze the performance degradation due to limited search budgets relative to ML performance, compare key metrics (worst-case and average complexity, hardware considerations), and highlight how advances in one approach transfer naturally to the other.
Our results clarify the operating regimes where GRAND and GCD demonstrate superior performance. This work provides both theoretical insights and practical guidelines for deploying universal guessing decoders in next-generation short-blocklength communications.
\end{abstract}


\section{Introduction}

With the full commercialization of 5G, research on beyond 5G (B5G) and 6G technologies is accelerating, bringing new challenges. B5G and 6G systems are expected to require extremely high data rates (up to Tbps), ultra-reliability (block error rates between \(10^{-5}\) and \(10^{-7}\)), and ultra-low latency (end-to-end delays of 0.1-1ms)~\cite{union2022future,rowshan2024channel}. To meet these stringent demands, short-blocklength channel codes and, in particular, efficient decoding algorithms for short codes have attracted significant attention~\cite{shirvanimoghaddam2018short}.

The emerging 6G ecosystem will span an even wider range of application scenarios—ultra-reliable low-latency control links, massive IoT, and more—each with distinct latency, reliability, and complexity requirements.  To support this diversity without proliferating a separate decoder for every code and use-case, there is growing interest in \emph{universal} decoding algorithms, which can flexibly accommodate any code structure with minimal customization~\cite{yue2023efficient}.

\begin{figure*}[tp]
  \centering
  \resizebox{0.60\textwidth}{!}{\begin{tikzpicture}[
    dot/.style = {circle, fill=black, inner sep=1pt, minimum size=6pt},
    every node/.style = {font=\large}
]
  \draw[-{Latex},thick] (0,0) -- (18,0);

  \node[dot, label=above:{Chase decoding\cite{1972Chase}}] at  (2,0) {};
  \node[dot, label=below:{Dorsch algorithm\cite{dorsch1974decoding}}] at  (5,0) {};
  \node[dot, label=above:{OSD\cite{fossorier1995soft}}] at  (8,0) {};
  \node[dot, label=below:{GRAND\cite{duffy2019capacity}}] at (11,0) {};
  \node[dot, label=above:{ORBGRAND\cite{duffy2022ordered}}] at (14,0) {};
  \node[dot, label=below:{GCD\cite{Ma2024}}] at (17,0) {};
\end{tikzpicture}}
  \caption{Milestones in the development of guessing decoding for short codes. The timeline highlights key algorithmic advances from Chase decoding to modern GRAND and GCD variants.}
  \label{fig:timeaxis}
\end{figure*}
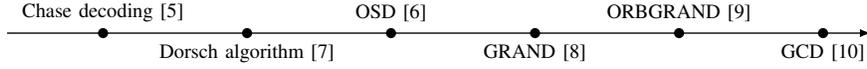

Guessing decoding is a representative universal and highly effective class of algorithms for short codes.
A timeline of key developments in the guessing decoding is illustrated in Fig.~\ref{fig:timeaxis}.
Early representatives include Chase decoding~\cite{1972Chase} and ordered statistics decoding~(OSD)~\cite{fossorier1995soft}.
Precisely, typical Chase-II decoding tests all combinations of flips on the least reliable bits (LRBs) based on the Hamming weight, followed by algebraic decoding~\cite{1972Chase}.
By contrast, most reliable basis~(MRB) reprocessing algorithms, such as Dorsch's algorithm~\cite{dorsch1974decoding} and the OSD algorithm~\cite{fossorier1995soft}, apply Gaussian elimination~(GE) to identify the MRB, then generate candidate codewords by re-encoding based on flipping patterns over the MRB, and finally select the highest-likelihood candidate as the decoding output. In Dorsch's algorithm, flipping patterns are generated according to the soft reliabilities, whereas in OSD, they are generated in increasing Hamming weight.

Inspired by practical demands on short codes, guessing-decoding algorithms such as guessing random additive noise decoding (GRAND)~\cite{duffy2018guessing,duffy2019capacity} and guessing codeword decoding~(GCD)~\cite{Ma2024,Zheng2024IT} have recently drawn significant interest. Unlike Chase or OSD, which typically rely on Hamming-weight-based patterns, GRAND and GCD generate flipping patterns ordered by soft reliability. Below we summarize their core ideas, key developments, and distinctions from classical guessing decoders.

The GRAND concept was formalized by Duffy, Li and M\'edard~\cite{duffy2018guessing,duffy2019capacity}.
It guesses error patterns over the entire received sequence in descending order of likelihood and performs simple code-membership checks.
In particular, it has been proven that GRAND is maximum-likelihood decoding~(MLD)~\cite{duffy2018guessing}. Subsequent important extensions include:
\begin{itemize}
  \item \emph{Flip-pattern strategies:} SRGRAND (symbol reliability)~\cite{duffy2021guessing}, DSGRAND (quantized reliability)~\cite{yuan2023guessing}, and the breakthrough integer-based ORBGRAND~\cite{duffy2022ordered}, which exploits the linear relationship between soft reliabilities and their rank orders to deliver a highly efficient implementation in hardware.
      Moreover, a theoretical analysis shows that ORBGRAND with random codebooks nearly attains the channel capacity~\cite{Mengxiao2023}.
  \item \emph{Constraint-based search reduction:} partially constrained-GRAND~(PC-GRAND)~\cite{wang2024partially}, constrained GRAND~\cite{rowshan2022constrained}, segmented GRAND~\cite{rowshan2025segmented}, and ordered reliability direct error pattern testing~\cite{Hadavian2025} and syndrome enhanced GRAND~\cite{rapp2025sogrand}, which leverages syndrome information to find candidate codewords early.
  \item \emph{Code-specific optimizations:} GRAND for 5G polar codes~\cite{duffy20205g},  and for BCH codes~\cite{Wang2024twostage}, which perform GRAND at the first stage and algebraic decoding instead of code-membership check at the second stage.
  \item \emph{Channel-specific optimizations:} GRAND for ISI channels~\cite{Ken2023,Duffy2025_ORBGRAND_AI_ISI},  impulsive noise models~\cite{feng2024laplacian}, and for fading channels using pseudo-soft information~\cite{Sarieddeen2022}.
  \item \emph{Applications:} for product codes with high-rate component codes~\cite{yuan2025soft,Riaz2025,Peng2024,Rapp2025}, for network-coding~\cite{su2024grand}, for demodulation or detection~\cite{Ozaydin2025,Chao2025}, and for multi-access scenarios~\cite{solomon2021managing,yang2023multiuser,Yang2024,yuan2023time}.
  \item \emph{Hardware implementations:} VLSI architectures~\cite{abbas2022high,condo2022fixed,Abbas2025Step} and the first ORBGRAND silicon in 40 nm CMOS, achieving 0.76 pJ/bit at 4.9 mW~\cite{riaz2024sub}.
\end{itemize}

\begin{table*}[ht]
  \centering
  \caption{Key differences between GRAND and GCD}
  \begin{tabular}{lcccc}
    \toprule
    & Pattern length & Re-encoding required? & Test sequence validity & Early termination criterion \\
    \midrule
    GRAND & \(n\) (entire received vector) & No                     & Often invalid, check parity & Stop at first valid codeword \\
    GCD   & \(k\) (partial received vector)        & Yes                    & Always valid (via re-encode) & \makecell{Stop if the ongoing guess is provable to be heavier} \\
    \bottomrule
  \end{tabular}
  \label{tab:grand-gcd}
\end{table*}

The GCD framework was formalized by Ma~\cite{Ma2024} as a universal decoder that flips bits in the encoding basis according to soft reliability and re-encodes to produce codeword candidates.
Unlike OSD, which requires GE to identify the MRB and generates flipping patterns in increasing Hamming weight order, GCD avoids GE entirely.
In particular, it has been proven that if the guesses are ordered by descending likelihood and an appropriate early stopping rule is applied, GCD is a non-exhaustive MLD algorithm~\cite{Zheng2024IT}.
Subsequent important extensions include:
\begin{itemize}
  \item \emph{Flip-order strategies:} ORB-GCD based on ordered reliability bits~\cite{Wang2024ordered}; enhanced GCD through ORBGRAND-AI~\cite{Feng2025EnhancedGCD}; locally-constrained GCD (LC-GCD)~\cite{Zheng2024ITW} and LC-OSD~\cite{Wang2022LCOSD}, which skip invalid patterns via partial constraints; GCD with single parity-check for reducing the guessing~\cite{griffin2025using}.
  \item \emph{Code-specific designs:} GCD as subdecoder for polar codes~\cite{Zheng2024IT}; GCD-like decoding for polar codes with transformed generator matrices~\cite{wang2024representative,Wang2024VTC}; GCD or OSD for LDPC codes~\cite{Wang2025ICCC}, cyclic-basis GCD for BCH codes without GE, approaching finite-length bounds~\cite{wang2024bch,wang2025globecom}.
  \item \emph{Applications:} soft-in/soft-out GCD for product codes~\cite{Feng2025GRAND,duffy2024soft,Peng2025}, GCD for quantum error-correction codes~\cite{Liang2025_BP_OSD_qLDPC,Liang2024_BP_LCOSD_Toric,Liang2024_Surface_List_QECC,Liang2025_QECC_Polar_APWDSIT}, LC-OSD in joint source-channel coding~\cite{Wang2024TCOMJSCC,QuasiOSD,Chen2023WCSP}, strong-interference communications~\cite{wang2024rateless} and short code constructions tailored for OSD-like algorithm~\cite{wang2024random,Wang2024ISIT,wang2023SGMC,zheng2025coding}, approaching capacity in finite-length region and achieving capacity in infinite-length region.
  \item \emph{Hardware implementation:} still largely unexplored, representing an open research direction.
\end{itemize}

The key distinctions between GRAND and GCD lie in: i) Pattern length: GRAND operates on flipping patterns over the entire received sequence with length~$n$, whereas GCD only applies flipping to the encoding basis of length~$k$; ii) Re-encoding requirement: GCD requires re-encoding to generate valid codeword candidates, while GRAND directly guesses each flipped sequence without re-encoding; iii) Validity of candidates: Most patterns tested in GRAND are invalid; in contrast, every candidate in GCD is generated via re-encoding and is therefore guaranteed to be a valid codeword. 
These distinctions are summarized in Table~\ref{tab:grand-gcd}.

\begin{figure}[tp]
  \centering
  \includegraphics[width=\figwidth]{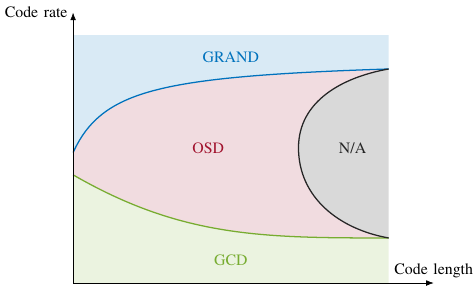}
  \caption{Practical operating regimes for three classes of guessing-based decoders over BPSK-AWGN channels. GRAND is best suited for high-rate codes, OSD performs well in the medium-rate regime due to GE, and GCD offers strong performance for low-rate codes.}
  \label{fig:Regions}
\end{figure}

While GRAND and GCD represent distinct strategies within guessing-decoding framework, they are not in competition. Indeed, they complement each other and are tailored to different code-rate regimes: GRAND is more effective at high rates, whereas GCD delivers superior performance at low rates. In contrast, OSD leverages GE to achieve robust mid-rate decoding, albeit with increased implementation complexity by GE. The preferred operating regions for these three strategies are illustrated in Fig.~\ref{fig:Regions}.
Moreover, GRAND and GCD benefit from one another: innovations for one method can often be adapted to the other, yielding new variants and performance improvements/complexity reduction.


Motivated by demands of decoding for short codes, this paper presents a unified and in-depth study of GRAND and GCD. Our contributions are summarized below:

\begin{enumerate}
  \item A systematic overview of various flip-pattern generation orders and implementations, including Hamming-weight, soft-information, and ORB-based strategies.
  \item Detailed algorithmic descriptions of GRAND and GCD, proofs of their ML optimality, saddle-point approximations for average queries, and numerical results.
  \item  Performance analysis of the gap to ML decoding, comparative study of key metrics, mutual improvements, and a survey of hardware implementations.
\end{enumerate}

The remainder of this paper is organized as follows.  Sec.~II covers preliminaries, including system models and flip-pattern generation.  Sec.~III and Sec.~IV present GRAND and GCD, respectively, including algorithm descriptions, ML-optimality proofs, saddle-point approximation for guesses on average, and numerical results.  Sec.~V compares the two decoders—performance gaps to ML, key metrics, cross-fertilization of techniques, and hardware implementations.  Sec.~VI concludes and outlines future directions.

\section{Preliminaries}

\subsection{System Model}
Let \(\mathbb{F}_2 = \{0,1\}\) denote the binary field.
Let \(\mathscr{C}[n,k]\) be a binary linear block code of length \(n\) and dimension \(k\), with parity-check matrix \(\mathbf{H}\) and generator matrix \(\mathbf{G}\).
An information vector \(\bm{u} \in \mathbb{F}_2^k\) is encoded into a codeword \(\bm{c} \in \mathbb{F}_2^n\) such that
$\mathbf{H} \bm{c}^\mathsf{T} = \mathbf{0}.$
The codeword \(\bm{c}\) is transmitted over a discrete memoryless channel~(DMC), yielding the received vector \(\bm{y} = (y_1, y_2, \dots, y_n)\).
Given $\bm{y}$, the log-likelihood ratio (LLR) of the \(i\)-th bit is defined as
\begin{equation}
\lambda_i = \ln \frac{P_{Y|C}(y_i \mid c_i = 0)}{P_{Y|C}(y_i \mid c_i = 1)}, \quad i = 1, 2, \dots, n,
\end{equation}
and the corresponding hard-decision vector \(\bm{z} \in \mathbb{F}_2^n\) is given by $z_i = 0$ if $\lambda_i \geq 0$ and $z_i = 1$ otherwise.
We refer to \(\lvert \lambda_i \rvert\) as the reliability of the $i$-th bit.

ML decoding seeks to find the codeword \( \bm{v}^* \) such that
\begin{equation}\label{Eq:MLdecoding}
\bm{v}^* = \underset{\bm{v} \in \mathscr{C}}{\arg\max} P_{Y | C}(\bm{y} | \bm{v}) = \underset{\bm{v} \in \mathscr{C}}{\arg\min} \ln \frac{P_{Y | C}(\bm{y} | \bm{z})}{P_{Y | C}(\bm{y} | \bm{v})}.
\end{equation}
Let \(\bm{e} = (e_1, \dots, e_n)\) be a test error pattern (TEP), then the corresponding candidate codeword is \(\bm{v} = \bm{z} \oplus \bm{e}\). The ML decoding problem becomes:
\begin{equation}
\bm{e}^* = \underset{\bm{e} \in \mathbb{F}_2^n,\, (\bm{z} \oplus \bm{e})\mathbf{H}^\mathsf{T} = \mathbf{0}}{\arg\min} \Gamma_{\mathrm{S}}(\bm{e}),
\end{equation}
where \(\Gamma_{\mathrm{S}}(\bm{e})\) is the \emph{soft weight} defined as
\begin{equation}\label{Eq:Gamma_S}
\Gamma_{\mathrm{S}}(\bm{e}) = \sum_{i=1}^n e_i \cdot \lvert \lambda_i \rvert.
\end{equation}

\subsection{Flipping Pattern Generation}
In both GRAND and GCD frameworks, the decoding performance and complexity depend heavily on the order in which TEPs are generated and tested.
In the following, we take TEPs of length \(n\) as an example to present a unified treatment of TEP-generation orders, applicable to both GRAND and GCD.
Three commonly used and typical TEP generation orders are reviewed, including Hamming-weight order, soft-weight order, and the ordered reliability bits (ORB) method.

\subsubsection{Hamming-weight order}
In classical algorithms such as Chase and OSD, TEPs are sorted by Hamming weights:
\begin{equation}\label{Eq:Hamming-weight}
  \Gamma_{\mathrm{H}}(\bm{e}^{(1)}) \leq \Gamma_{\mathrm{H}}(\bm{e}^{(2)}) \leq \cdots,
\end{equation}
where $\Gamma_{\mathrm{H}}(\cdot)$ is the Hamming weight of a sequence, i.e., $\Gamma_{\mathrm{H}}(\bm{e}) = \sum_{i=1}^{n} e_i $.

\subsubsection{Soft-weight order}
This order is essentially equivalent to likelihood order, where TEPs are sorted by
\begin{equation}\label{Eq:soft-order}
\Gamma_{\mathrm{S}}({\bm{e}}^{(1)}) \leq \Gamma_{\mathrm{S}}({\bm{e}}^{(2)}) \leq \cdots,
\end{equation}
where $\Gamma_{\mathrm{S}}(\cdot)$ is defined as~\eqref{Eq:Gamma_S}.
This order is optimal in terms of ML performance, but naive enumeration is computationally infeasible. A structured solution is to organize all TEPs into a tree, such as the flipping pattern tree~(FPT)~\cite{tang2022new}, which specifes a partial order \(\preceq\) between patterns. Specifically, for \(\bm{e}, \bm{e}' \in \mathbb{F}_2^n\), we write \(\bm{e} \preceq \bm{e}'\) if there exists an injection \(\pi: \text{supp}(\bm{e}) \to \text{supp}(\bm{e}')\) such that \(i \leq \pi(i)\) for all \(i\). Then, assuming that the LLR magnitudes are sorted: \(\lvert \lambda_1 \rvert \leq \lvert \lambda_2 \rvert \leq \cdots \leq \lvert \lambda_n \rvert\), we have
$
\bm{e} \preceq \bm{e}' ~ \text{implying}~ \Gamma_{\mathrm{S}}(\bm{e}) \leq \Gamma_{\mathrm{S}}(\bm{e}').
$
By the FPT, TEPs can be efficiently enumerated~\cite{tang2022new}. Similar trees for binary codes to generate the TEPs were also constructed in GRAND~\cite[Algorithm~2]{solomon2020soft}.

\subsubsection{ORB Order}

The ORB method maps soft reliabilities to integer ranks, enabling hardware-efficient TEP enumeration by integer partition~\cite{duffy2022ordered}.
This relies on the empirical observation that sorted LLR magnitudes exhibit an approximately linear relationship with their rank~\cite{duffy2022ordered}.
Formally, let \(\Lambda_i\in\{1,\dots,n\}\) denote the rank of bit \(i\) when \(\lvert\lambda_i\rvert\) are sorted in descending order.  The \emph{logistic weight} of a TEP \(\bm{e}\) is then
\begin{equation}\label{eq:logistic_weight}
\Gamma_{\mathrm{L}}(\bm{e}) =\sum_{i=1}^{n} \Lambda_i\,e_i.
\end{equation}
As a result, TEPs can be tested in non-decreasing logical weight as
\begin{equation}\label{eq:orb_order}
\Gamma_{\mathrm{L}}(\bm{e}^{(1)}) \le \Gamma_{\mathrm{L}}(\bm{e}^{(2)}) \le \cdots.
\end{equation}
In particular, an integer partition pattern generator can be used to generate the TEPs for a given logistic weight, ensuring the order of \eqref{eq:orb_order}.

For example, consider the case where $n=4$ and assume $\lambda_1 \leq \lambda_2 \leq \lambda_3 \leq \lambda_4$.
The integers are decomposed to generate TEPs with increasing logistic weight:
\begin{itemize}
  \item \(1 \to\) \texttt{1000}
  \item \(2 \to\) \texttt{0100}
  \item \(3 \to\) \texttt{0010}, \texttt{1100}
\end{itemize}

\begin{example}[Comparison of Ordering Strategies]
Consider $n = 4$ and  \(\bm{\lambda} = (2, 3, 4, 8)\).
Table~\ref{Table:order} lists the Hamming weight \(\Gamma_{\mathrm{H}}(\bm{e})\), soft weight \(\Gamma_{\mathrm{S}}(\bm{e})\), and logistic weight \(\Gamma_{\mathrm{L}}(\bm{e})\) for patterns up to Hamming weight of two.
The partial test orders of these three metrics are:
\\0000, 1000, 0100, 0010, 0001, 1100, … (soft-weight);
\\0000, 1000, 0100, 0010, 1100, 1010, … (Hamming-weight);
\\0000, 1000, 0100, 0010, 1100, 0001, … (logistic-weight).

\begin{table}[!t]
\centering
\caption{Ordering of TEPs for \(n=4\) under three metrics.}
\label{Table:order}
\begin{tabular}{|c|c|c|c|}
\hline
\textbf{TEP} & \(\Gamma_{\mathrm{H}}(\bm{e})\) & \(\Gamma_{\mathrm{S}}(\bm{e})\) & \(\Gamma_{\mathrm{L}}(\bm{e})\) \\
\hline
0000 & 0 & 0  & 0 \\
1000 & 1 & 2  & 1 \\
0100 & 1 & 3  & 2 \\
0010 & 1 & 4  & 3 \\
0001 & 1 & 8  & 4 \\
\hline
1100 & 2 & 5  & 3 \\
1010 & 2 & 6  & 4 \\
0110 & 2 & 7  & 5 \\
1001 & 2 & 10 & 5 \\
0101 & 2 & 11 & 6 \\
0011 & 2 & 12 & 7 \\
\hline
\end{tabular}
\end{table}
\end{example}

\textbf{\emph{Remark.}}
Hamming-weight ordering is simple but does not fully exploit soft information, resulting in suboptimal performance with limited search budgets. Soft-weight ordering is ML-optimal but difficult to implement in real time. ORB ordering offers a practical compromise: its integer-based rank approximation is hardware-friendly, can be precomputed or implemented with minimal logic overhead, and achieves near-soft-weight performance with high VLSI efficiency.

\section{Guessing-Noise-Based Decoder}

In this section, we introduce the guessing-noise-based decoder, commonly known as GRAND.  We begin with a description of the algorithm’s implementation, then provide a concise proof of its ML optimality.  Next, we derive a saddle-point approximation for the average number of guesses required by GRAND.  Finally, we present numerical results illustrating its performance.

\medskip
\noindent\textbf{Complexity scaling with redundancy.}
It has been shown that, for any \([n,k]\) code, the number of queries until the first incorrect codeword hit is (approximately) geometrically distributed with mean \(2^{\,n-k}\)~\cite[Theorem~2]{duffy2019capacity}.  Since GRAND terminates immediately upon the first valid codeword, an upper bound on its average complexity depends essentially on the number of redundant bits \(n-k\), rather than on \(n\) or the rate \(k/n\) alone.  This makes GRAND especially attractive for moderate-redundancy codes of arbitrary length.
Recent work~\cite{liang2025guesswork} shows that this expected complexity is conservative: the average number of queries is approximately $2^{n-k} \varepsilon_{\textrm{RCU}}$, with $\varepsilon_{\textrm{RCU}}$
denoting the RCU bound for codes of length $n$ and dimension $k$.
See Sec.~\ref{Subsection:KeyMetricComparison} for a detailed derivation.
This reinforces that GRAND is especially efficient in high-SNR scenarios, where very low error rates are sought.

\subsection{Algorithmic Implementation}

Given a TEP generator that produces the sequence \(\{\bm{e}^{(i)}\}_{i=1}^{2^n}\) in a prescribed order—e.g., the soft-weight order of \eqref{Eq:soft-order} or the ORB order of \eqref{eq:orb_order}—GRAND tests each \(\bm{e}^{(i)}\) in turn until a valid pattern is found or the maximum query limit is reached.
Formally, for each \(\bm{e}^{(\ell)}\), check whether
\[
  \bm{e}^{(\ell)}\mathbf{H}^\mathsf{T} = \bm{s},
  \quad\text{where }\bm{s} = \bm{z}\,\mathbf{H}^\mathsf{T}.
\]
If this holds, output \(\hat{\bm{c}}=\bm{z} \oplus \bm{e}^{(\ell)}\) and terminate.
For the sake of clarity, GRAND algorithm is summarized in Algorithm~\ref{alg:GRAND}.

\begin{algorithm}[!t]
  \caption{GRAND Algorithm}
    \renewcommand{\algorithmicrequire}{\textbf{Input:}}
  \renewcommand{\algorithmicensure}{\textbf{Output:}}
  \label{alg:GRAND}
  \begin{algorithmic}[1]
    \Require Parity-check matrix \(\mathbf{H}\in\mathbb{F}_2^{(n-k)\times n}\), LLR vector \(\bm{\lambda}\in\mathbb{R}^n\), max number of guesses \(\ell_{\max}\), hard-decision sequence $\bm z$, flipping-pattern order
    \Ensure Decoded codeword \(\hat{\bm{c}}\in\mathbb{F}_2^n\)

    \Function{GRAND}{$\mathbf{H},\,\bm{\lambda},\,\bm{z},\,\ell_{\max},\,\text{order}$}
    \State Compute syndrome \(\bm{s}\gets \bm{z}\,\mathbf{H}^\mathsf{T}\)
    \State Initialize \(\hat{\bm{c}}\gets \bm{z}\) 
      \For{$\ell \gets 1$ \textbf{to} $\ell_{\max}$}
        \State Generate \(\bm{e}^{(\ell)}\) according to the specified order
        \If{$\bm{e}^{(\ell)}\mathbf{H}^\mathsf{T} = \bm{s}$}
          \State \(\hat{\bm{c}}\gets \bm{z} \oplus \bm{e}^{(\ell)}\)
          \State  \textbf{break}
        \EndIf
      \EndFor
      \State \Return \(\hat{\bm{c}}\)
    \EndFunction

  \end{algorithmic}
\end{algorithm}

\subsection{Optimality Condition}

\begin{definition}[Membership-Check Stopping Criterion]\label{def:trivial_GRAND}
Terminate GRAND whenever
\[
\bm{e}^{(\ell)}\mathbf{H}^\mathsf{T}=\bm{s},
\]
and output \(\hat{\bm{c}}=\bm{z} \oplus \bm{e}^{(\ell)}\).
\end{definition}

\begin{proposition}\label{prop:trivial-ml-GRAND}
If TEPs are generated in non-decreasing soft-weight order as in~\eqref{Eq:soft-order}, then the first codeword output by GRAND under the membership-check stopping criterion is an ML codeword.
\end{proposition}

\begin{IEEEproof}
At iteration \(j\):
\[
\bm{e}^{(j)}\mathbf{H}^\mathsf{T}=\bm{s}.
\]
By the assumption of soft-weight order~\eqref{Eq:soft-order}, \(\Gamma_{\mathrm{S}}(\bm{e}^{(j)})\le\Gamma_{\mathrm{S}}(\bm{e}^{(\ell)})\) for all \(\ell>j\), no subsequent TEP can have lower soft weight.  Hence \(\bm{e}^{(j)}\) yields an ML codeword.
\end{IEEEproof}

\subsection{Average Number of Queries}\label{subseciton:GRAND-queries}

In this subsection, we estimate the average number of queries required by GRAND using a saddle-point approximation.  Consider transmission over a binary-input output-symmetric (BIOS) channel and, without loss of generality, assume that the all-zero codeword is transmitted.  Let \(\bm{e}^*\) be the lightest valid TEP (the ML pattern) under soft-weight ordering \eqref{Eq:soft-order}.  Define
\begin{equation}
  \mathcal{L}_{\mathrm{N}}^*(\bm{\lambda})
  = \bigl\{\bm{f}\in\mathbb{F}_2^n \!:\! \Gamma_{\mathrm{S}}(\bm{f}) \!\le\! \Gamma_{\mathrm{S}}(\bm{e}^*) |\bm{\lambda} \bigr\},
  L_{\mathrm{N}}^*(\bm{\lambda})
  = \bigl|\mathcal{L}_{\mathrm{N}}^*(\bm{\lambda})\bigr|.
\end{equation}
Since \(\bm{e}^*\) depends on both \(\bm{\lambda}\) and the code, direct evaluation of \(L_{\mathrm{N}}^*(\bm{\lambda})\) is difficult.  Instead, let \(\bm{e}=\bm{z}\) be the true TEP when the zero codeword is sent, and consider
\begin{equation}
  \mathcal{L}_{\mathrm{N}}(\bm{\lambda})
  = \bigl\{\bm{f}\in\mathbb{F}_2^n \!:\! \Gamma_{\mathrm{S}}(\bm{f}) \!\le\! \Gamma_{\mathrm{S}}(\bm{e}) |\bm{\lambda} \bigr\},
  L_{\mathrm{N}}(\bm{\lambda})
  = \bigl|\mathcal{L}_{\mathrm{N}}(\bm{\lambda})\bigr|.
\end{equation}
Since \(\Gamma_{\mathrm{S}}(\bm{e}^*)\le\Gamma_{\mathrm{S}}(\bm{e})\), we have \(\mathcal{L}_{\mathrm{N}}^*\subseteq\mathcal{L}_{\mathrm{N}}\) and hence
\(L_{\mathrm{N}}^*(\bm{\lambda})\le L_{\mathrm{N}}(\bm{\lambda})\). Clearly, \(L_{\mathrm{N}}(\bm{\lambda})\) is easier to estimate.
To compute \(L_{\mathrm{N}}(\bm{\lambda})\), we adopt a uniform distribution over \(\mathbb{F}_2^n\):
\begin{equation}
  \mathbb{P}[\bm{f}=\bm{f}^*]=2^{-n},\quad\forall\,\bm{f}^*\in\mathbb{F}_2^n.
\end{equation}
Under this model, the list size is
\begin{align}
  {L}_{\textrm{N}}(\bm{\lambda})
  \triangleq 2^{n} \mathbb{P}\Biggl[\left. \sum_{i=1}^{n} (f_i - e_i) |\lambda_i| \leq 0 \right|\bm{\lambda}\Biggr]
  = 2^{n} \mathbb{P}[ S_{n} \leq 0 |\bm{\lambda}],
\end{align}
where $S_{n} = \sum_{i=1}^{n} D_i$ and $D_i = (f_i - e_i)\,|\lambda_i|\in\{0,|\lambda_i|\}$ each with probability $1/2$.
Therefore, calculating ${L}_{\textrm{N}}(\bm{\lambda})$ is equivalent to evaluating the tail probability, which allows us to use the saddlepoint approximation~\cite{Font2018Saddlepoint}.
That is, ${L}_{\textrm{N}}(\bm{\lambda})$ can be approximated by
\begin{equation}
  \label{eq:GRAND-sd}
 {L}_{\textrm{N}}(\bm{\lambda})\! \approx \! 2^{n-1} \exp\biggl(\!{\kappa(\hat{s})\!+ \!\frac{\hat{s}^2\kappa''(\hat{s})}{2}}\!\biggr)
  {\rm erfc} \biggl(- \hat{s}\sqrt{\frac{\kappa''(\hat{s})}{2}}\biggr),
\end{equation}
where ${\rm erfc}(\cdot)$ is the complementary error function and other functions can be found in~\cite{Font2018Saddlepoint}.

At this point, the distribution of \( L_{\textrm{N}}(\bm{\lambda}) \) can be estimated through Monte Carlo simulations by sampling a sufficiently large number of  received sequences (corresponding to the LLR sequences \( \bm{\lambda} \)).
Averaging this expression over the distribution of \(\bm{\lambda}\) yields the mean query count \(\overline{L}_{\mathrm{N}} =  \mathbb{E}\bigl[{L}_{\textrm{N}}(\bm{\lambda})\bigr]\).
Likewise, the tail probability \(\mathbb{P}[L_{\mathrm{N}}>\ell_{\max}]\) can be obtained by evaluating the corresponding saddle-point ratio.

\textbf{\emph{Remark.}}
Although we employ simulated LLR samples to evaluate the saddle-point expressions, this method is agnostic to the code’s structure and does not require  decoding.  As a result, it provides rapid, reliable estimates of GRAND's complexity for any \([n,k]\) code over a given channel.

\subsection{Numerical Results}
\begin{figure}[tp]
  \centering
  \subfloat[FER \label{fig:QOSD-G-fer}]{\includegraphics[width=0.22\textwidth]{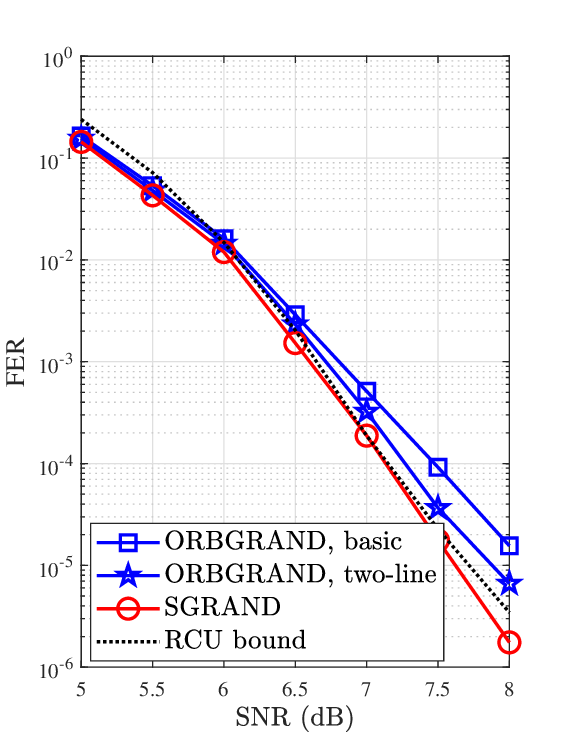}}
  \subfloat[The average number of queries]{\includegraphics[width=0.22\textwidth]{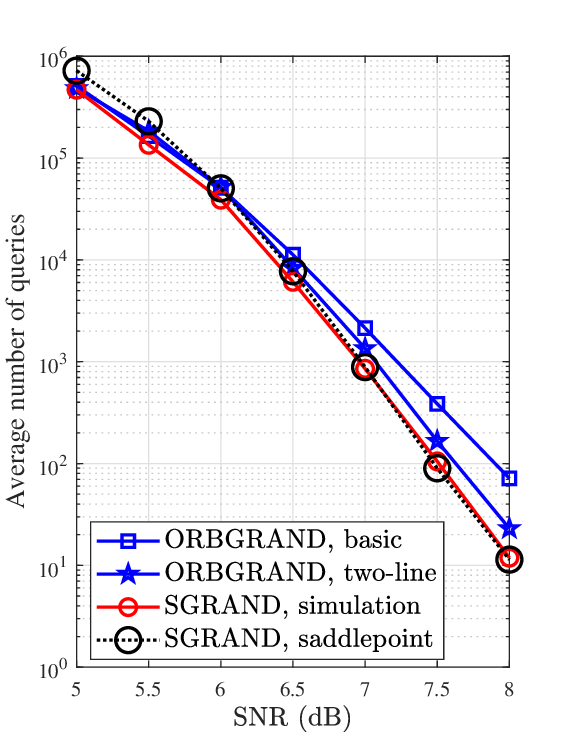}}
  \caption{
       The performance of the random linear code $\mathscr{C}[128,106]$ under SGRAND~\cite{duffy2019capacity} and ORBGRAND~\cite{duffy2022ordered}, where $\ell_{\max} = 5\times 10^6$.}
    \label{fig:ORBGRAND_SGRAND}
\end{figure}

\begin{example}[GRAND with Soft-Weight Order versus ORB Order]\label{Example:decoding-time}
Consider the random linear code $\mathscr{C}[128, 106]$ over  BPSK-AWGN channels.
We simulate GRAND with $\ell_{\max} = 5\times 10^6$ under two different TEP orders:
(1) the soft-weight order as \eqref{Eq:soft-order}, referred to as SGRAND~\cite{duffy2019capacity};
(2) the ORB order as \eqref{eq:orb_order}, referred to as ORBGRAND~\cite{duffy2022ordered}.
The FER is shown in Fig.~\ref{fig:ORBGRAND_SGRAND}(a), from which we observe that both SGRAND and ORBGRAND can closely approach the RCU bound.
Moreover, ORBGRAND’s performance closely matches that of SGRAND, highlighting the effectiveness of the ORB approximation.
The average number of queries is shown in Fig.~\ref{fig:ORBGRAND_SGRAND}(b), from which we observe that the simulated results closely match the proposed saddle-point approximation, confirming the accuracy and usefulness of the theoretical analysis.
\end{example}

\section{Guessing-codeword-based Decoder}
In this section, we introduce the guessing-codeword-based decoder, abbreviated as GCD. We begin with a description of the algorithm’s implementation, then provide a proof of its ML optimality. Next, we derive a saddle-point approximation for the average number of guesses required by GCD. In the final part of this section, we present numerical results illustrating its performance.

\subsection{Algorithmic Implementation}

Without loss of generality, we assume that the last $(n - k)$ columns of the parity-check matrix $\mathbf{H}$ are linearly independent. In this case, $\mathbf{H}$ can be transformed into a systematic form through elementary row operations:
\begin{equation}
  \label{equ:systematic}
\mathbf{H} \;\rightarrow\;
\begin{bmatrix}
  \mathbf{P} & \mathbf{I}
\end{bmatrix},
\end{equation}
where $\mathbf{P}$ is an $(n - k) \times k$ sub-matrix and $\mathbf{I}$ is the identity matrix of size $n - k$.
Accordingly, any TEP \(\bm{e}\in\mathbb{F}_2^n\) can be partitioned as
$
\bm{e} = (\,\bm{e}_{\mathrm{I}},\,\bm{e}_{\mathrm{P}}\,),
$
where \(\bm{e}_{\mathrm{I}}\in\mathbb{F}_2^k\) covers the information bits and \(\bm{e}_{\mathrm{P}}\in\mathbb{F}_2^{\,n-k}\) covers the parity bits.  For a valid TEP, the parity part is uniquely determined by
\[
\bm{e}_{\mathrm{P}} = \bm{s} - \bm{e}_{\mathrm{I}}\,\mathbf{P}^{\mathsf{T}},
\]
where \(\bm{s} = \bm{z}\,\mathbf{H}^{\mathsf{T}}\) is the syndrome of the hard-decision vector.

GCD generates information-part patterns in a specific order, such as non-decreasing soft-weight order of~\eqref{Eq:soft-order} as
\begin{equation}\label{eq:GCD_sorting}
\Gamma_{\mathrm{S}}\bigl(\bm{e}_{\mathrm{I}}^{(1)}\bigr)
\le
\Gamma_{\mathrm{S}}\bigl(\bm{e}_{\mathrm{I}}^{(2)}\bigr)
\le
\cdots
\le
\Gamma_{\mathrm{S}}\bigl(\bm{e}_{\mathrm{I}}^{(2^k)}\bigr),
\end{equation}
where \(\Gamma_{\mathrm{S}}(\bm{e}_{\mathrm{I}})\) is the soft weight of the information positions.

\begin{algorithm}[!t]
  \caption{GCD Algorithm}
    \renewcommand{\algorithmicrequire}{\textbf{Input:}}
  \renewcommand{\algorithmicensure}{\textbf{Output:}}
  \label{alg:GCD}
  \begin{algorithmic}[1]
    \Require Systematic parity-check matrix $[\mathbf{P}, \mathbf{I}]\in\mathbb{F}_2^{(n-k)\times n}$,
             LLR vector $\bm{\lambda}\in\mathbb{R}^n$, max number of guesses $\ell_{\max}$, hard-decision sequence $\bm z$, flipping-pattern order, optional stopping criterion $\Psi$
    \Ensure Decoded codeword $\hat{\bm{c}}\in\mathbb{F}_2^n$

    \Function{GCD}{$[\mathbf{P}, \mathbf{I}],\,\bm{z},\,\bm{\lambda},\,\ell_{\max},\,\text{order},\,\Psi$}
    \State Compute syndrome $\bm{s}\gets \bm{z}\,[\mathbf{P}, \mathbf{I}]^\mathsf{T}$
    \State Initialize $\bm{e}_{\mathrm{opt}}\gets (\bm{0}^k,\bm{s})$, $\Gamma_{\mathrm{opt}}\gets +\infty$
      \For{$\ell \gets 1$ \textbf{to} $\ell_{\max}$}
        \State Generate $\bm{e}_{\mathrm{I}}^{(\ell)}$ according to the specified order
        \If{\(\Psi\) is defined and \(\Psi(\bm{e}^{(\ell)})\) holds}
          \State \textbf{break}
        \EndIf
        \State $\bm{e}_{\mathrm{P}}^{(\ell)} \gets \bm{s} - \bm{e}_{\mathrm{I}}^{(\ell)}\mathbf{P}^\mathsf{T}$
        \State $\bm{e}^{(\ell)} \gets (\bm{e}_{\mathrm{I}}^{(\ell)},\,\bm{e}_{\mathrm{P}}^{(\ell)})$
        \If{$\Gamma_{\mathrm{S}}(\bm{e}^{(\ell)}) < \Gamma_{\mathrm{opt}}$}
          \State $\bm{e}_{\mathrm{opt}}\gets \bm{e}^{(\ell)}$
          \State $\Gamma_{\mathrm{opt}}\gets \Gamma_{\mathrm{S}}(\bm{e}^{(\ell)})$
        \EndIf
      \EndFor
      \State \Return$\hat{\bm{c}}\gets \bm{z} \oplus \bm{e}_{\mathrm{opt}}$
    \EndFunction

  \end{algorithmic}
\end{algorithm}

Unlike GRAND, which stops at the first valid TEP, GCD stops by comparing the up-to-date optimal candidate and the newly produced TEP in terms of soft weight.
GCD algorithm with an optional early
stopping criterion (see Sec.~\ref{GCDSubsec:Stopping} for details) is summarized in Algorithm~\ref{alg:GCD}.

\subsection{Early Stopping Criteria and ML Optimality}~\label{GCDSubsec:Stopping}

In this subsection, we introduce two early stopping criteria for GCD algorithm.  First, we define a trivial stopping rule and prove that it preserves ML optimality.  Then, to further reduce the average number of guesses, we propose a dynamic approximate ideal (DAI) criterion that uses a statistical estimate of the parity-part soft weight.

Recall that in GCD algorithm we generate a sequence of TEPs
$
\bm{e}^{(\ell)} = \bigl(\bm{e}_{\mathrm{I}}^{(\ell)},\,\bm{e}_{\mathrm{P}}^{(\ell)}\bigr),
\quad \ell = 1,2,\dots,
$
where \(\bm{e}_{\mathrm{I}}^{(\ell)}\in\mathbb{F}_2^k\) is the information-part and \(\bm{e}_{\mathrm{P}}^{(\ell)}\in\mathbb{F}_2^{n-k}\) is the parity-part.  After the \(j\)-th re-encoding, define the optimal pattern so far by
\begin{equation}\label{eq:eopt}
\bm{e}_{\mathrm{opt}}^{(j)}
=\underset{\bm{e}^{(\ell)}: 1\le \ell\le j}{\arg\min}\Gamma_{\mathrm{S}}\bigl(\bm{e}^{(\ell)}\bigr).
\end{equation}

\begin{definition}[Trivial Stopping Criterion]\label{def:trivial}
Terminate the decoder whenever
\[
\Gamma_{\mathrm{S}}\bigl(\bm{e}_{\mathrm{opt}}^{(j)}\bigr)
\le
\Gamma_{\mathrm{S}}\bigl(\bm{e}_{\mathrm{I}}^{(j)}\bigr),
\]
and output
$
\hat{\bm{c}}
=
\bm{z} \oplus \bm{e}_{\mathrm{opt}}^{(j)}.
$
\end{definition}

\begin{proposition}\label{prop:trivial-ml-GCD}
If the trivial stopping criterion is met at the \(j\)-th guess, then the delivered codeword \(\hat{\bm{c}}\) is ML optimal.
\end{proposition}

\begin{IEEEproof}
Suppose at iteration \(j\) we have
\(\Gamma_{\mathrm{S}}(\bm{e}_{\mathrm{opt}}^{(j)}) \le \Gamma_{\mathrm{S}}(\bm{e}_{\mathrm{I}}^{(j)})\).
By the soft-weight ordering~\eqref{Eq:soft-order},
\(\Gamma_{\mathrm{S}}(\bm{e}_{\mathrm{I}}^{(j)}) \le \Gamma_{\mathrm{S}}(\bm{e}_{\mathrm{I}}^{(\ell)})\) for all \(\ell>j\).
Since \(\Gamma_{\mathrm{S}}(\bm{e}^{(\ell)}) = \Gamma_{\mathrm{S}}(\bm{e}_{\mathrm{I}}^{(\ell)}) + \Gamma_{\mathrm{S}}(\bm{e}_{\mathrm{P}}^{(\ell)}) \ge \Gamma_{\mathrm{S}}(\bm{e}_{\mathrm{I}}^{(\ell)})\),
we have
\[
\Gamma_{\mathrm{S}}(\bm{e}_{\mathrm{opt}}^{(j)})
\le
\Gamma_{\mathrm{S}}(\bm{e}^{(\ell)})
\quad\forall\,\ell>j.
\]
Hence, \(\bm{e}_{\mathrm{opt}}^{(j)}\) has the minimum soft weight over all patterns and yields the ML codeword.
\end{IEEEproof}

For low-rate codes, the trivial stopping criterion becomes ineffective in reducing search complexity because the partial TEP ${\bm{e}}_{\mathrm{I}}$ (corresponding to the information part) has a relatively short length $k$.
Therefore, we turn to the ideal stopping condition~\cite{liang2023randomarXiv} for GCD:
$
\Gamma_{\mathrm{opt}} \le
\Gamma_{\mathrm{S}}\bigl(\bm{e}_{\mathrm{I}}^{(\ell)}\bigr)
+\Gamma_{\mathrm{S}}\bigl(\bm{e}_{\mathrm{P}}^{(\ell)}\bigr).
$
As proven in~\cite{liang2023randomarXiv}, GCD-like decoding with this ideal condition is ML decoding. During the search process, $\Gamma_{\mathrm{S}}(\bm{e}_{\mathrm{I}}^{(\ell)})$ is computable since $\bm{e}_{\mathrm{I}}^{(\ell)}$ is explicitly hypothesized. However, $\Gamma_{\mathrm{S}}(\bm{e}_{\mathrm{P}})$ requires knowledge of the true $\bm{e}_{\mathrm{P}}$, which is unknown at the receiver. To address this, we estimate $\Gamma_{\mathrm{S}}(\bm{e}_{\mathrm{P}})$ using its statistical expectation as
\begin{equation}\label{Eq:tau}
   \tau =  \mathbb{E}\bigl[\Gamma_{\mathrm{S}}(\bm{e}_{\mathrm{P}}) \mid \bm{\lambda}\bigr] = \sum_{i=k +1}^{n} \frac{|{\lambda_i}|}{1+\exp(|{\lambda_i}|)}.
\end{equation}
Based on the above description, we present the DAI stopping criterion for the proposed algorithm as follows.

\begin{definition}[DAI Stopping Criterion]\label{def:dai}
Terminate the decoder whenever
\[
\Gamma_{\mathrm{S}}\bigl(\bm{e}_{\mathrm{opt}}^{(j)}\bigr)
\le
\Gamma_{\mathrm{S}}\bigl(\bm{e}_{\mathrm{I}}^{(j)}\bigr)
+\tau,
\]
where
\(\tau\) is given by \eqref{Eq:tau}. Then output
\(\hat{\bm{c}}
=
\bm{z} \oplus \bm{e}_{\mathrm{opt}}^{(j)}\).
\end{definition}

\textbf{\emph{Remark.}}
Although the DAI criterion does not guarantee ML optimality, incorporating the compensation term $\tau$ dramatically reduces the average number of guesses while maintaining negligible FER degradation, as numerically demonstrated in Section~III-D.

\subsection{Average Number of Queries}\label{subseciton:GCD-queries}

In this subsection, we derive approximations for the average number of queries required by GCD algorithm, both with the trivial stopping criterion and with the DAI stopping criterion.

\subsubsection{GCD with Trivial Stopping Criterion}

Let the true error pattern be \(\bm{e}=(\bm{e}_{\mathrm{I}},\bm{e}_{\mathrm{P}})\).  Under the trivial stop condition,
GCD should examine all partial patterns \(\bm{f}_{\mathrm{I}}\) whose soft weight does not exceed \(\Gamma_{\mathrm{S}}(\bm{e})\) to ensure the true error pattern \(\bm{e}\) is included.
 Define the corresponding search set
\begin{equation}\label{Eq:seachset}
  \mathcal{L}_{\mathrm{C}}(\bm{\lambda})
=
\bigl\{\bm{f}\in\mathbb{F}_2^n : \Gamma_{\mathrm{S}}(\bm{f}_{\mathrm{I}})\le \Gamma_{\mathrm{S}}(\bm{e}),\ \bm{f}_{\mathrm{P}}=\bm{0}|\bm{\lambda}\bigr\},
\end{equation}
and let
$
L_{\mathrm{C}}(\bm{\lambda})
= \bigl\lvert \mathcal{L}_{\mathrm{C}}(\bm{\lambda})\bigr\rvert.
$
Since \(\bm{f}_{\mathrm{I}}\) is uniformly distributed over \(\mathbb{F}_2^k\), we have
\begin{align*}
L_{\mathrm{C}}(\bm{\lambda})
&= 2^k \mathbb{P}\Bigl[\Gamma_{\mathrm{S}}(\bm{f}_{\mathrm{I}})\le \Gamma_{\mathrm{S}}(\bm{e})\Bigm|\bm{\lambda}\Bigr]\\
&= 2^k \mathbb{P}\Bigl[S_k \le \sum_{i=k+1}^n e_i\,\lvert\lambda_i\rvert\Bigm|\bm{\lambda}\Bigr],
\end{align*}
where \(S_k=\sum_{i=1}^k f_i\,\lvert\lambda_i\rvert\).
By a similar saddlepoint approximation as for GRAND, the distribution of the queries and the average number of searches
\( \overline{L}_{\textrm{C}} \triangleq \mathbb{E}\bigl[L_{\textrm{C}}(\bm{\lambda})\bigr] \) can be evaluated via Monte Carlo
simulations by sampling a large number LLR sequences.

\subsubsection{GCD with DAI Criterion}

When using the DAI stopping rule, the decoder terminates once
\(\Gamma_{\mathrm{S}}(\bm{e}_{\mathrm{I}})\,+\,\tau \le \Gamma_{\mathrm{S}}(\bm{e})\), where
\(\tau\) is given by \eqref{Eq:tau}.  The search set becomes
\[
\mathcal{L}_{\mathrm{DAI}}(\bm{\lambda})
=
\bigl\{\bm{f}\in\mathbb{F}_2^n : \Gamma_{\mathrm{S}}(\bm{f}_{\mathrm{I}}) + \tau \le \Gamma_{\mathrm{S}}(\bm{e}),\ \bm{f}_{\mathrm{P}}=\bm{0}|\bm{\lambda}\bigr\},
\]
with cardinality
\[
L_{\mathrm{DAI}}(\bm{\lambda})
= \bigl\lvert \mathcal{L}_{\mathrm{DAI}}(\bm{\lambda})\bigr\rvert
= 2^k\mathbb{P}\Bigl[S_k \le \sum_{i=k+1}^n e_i\,\lvert\lambda_i\rvert - \tau\Bigm|\bm{\lambda}\Bigr].
\]
Similarly, the distribution of the queries and the average number of searches
\( \overline{L}_{\textrm{DAI}} \triangleq \mathbb{E}\bigl[L_{\textrm{DAI}}(\bm{\lambda})\bigr] \) can be evaluated via Monte Carlo sampling of received LLR sequences.

\subsection{Numerical Results}

\begin{figure}[tp]
  \centering
  \subfloat[FER \label{fig:DAI_vs_Trivial_FER}]{\includegraphics[width=0.22\textwidth]{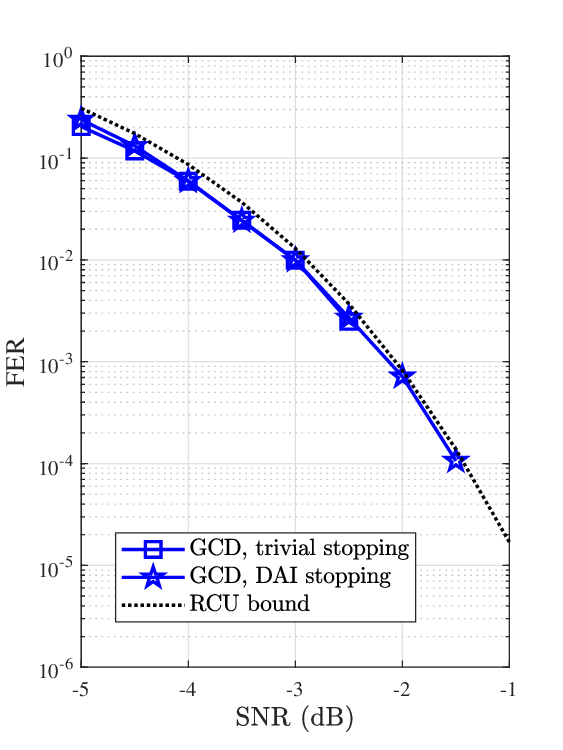}}
  \subfloat[Average number of queries \label{fig:DAI_vs_Trivial_Searches}]{\includegraphics[width=0.22\textwidth]{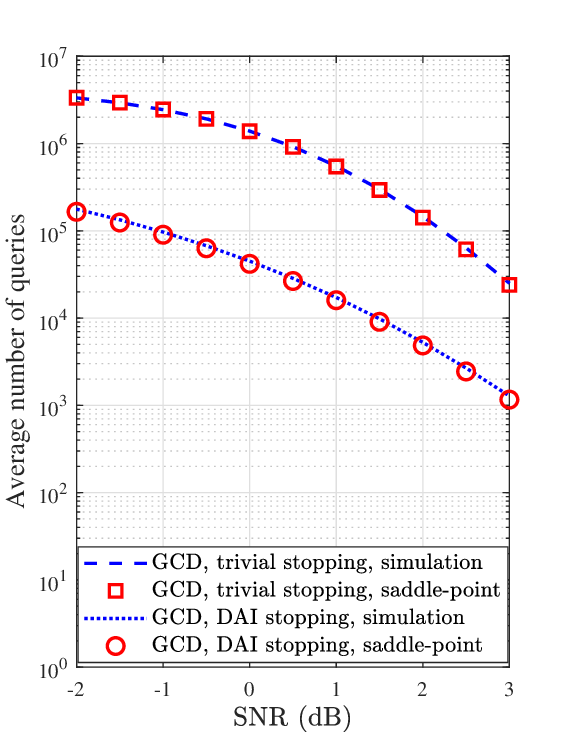}}
  \caption{The performance of the eBCH code $\mathscr{C}[128,22]$ under GCD with trivial and DAI stopping criteria, where  $\ell_{\max} = 5\times 10^6$.}
    \label{fig:DAI_vs_Trivial}
\end{figure}

\begin{example}[Trivial versus DAI Stopping Criterion]\label{Example:DAI_vs_Trivial}
Consider the  eBCH code \( \mathscr{C}[128,22] \) over BPSK-AWGN channels. GCD algorithm (with both trivial and DAI stopping criteria) is employed for decoding, where the maximum number of searches is \(\ell_{\max} = 5\times 10^6\).
The FER results are shown in Fig.~\ref{fig:DAI_vs_Trivial}(a), from which we observe that GCD algorithm with DAI stopping is nearly identical to that with trivial stopping, both approaching the RCU bound.
The average number of queries is shown in Fig.~\ref{fig:DAI_vs_Trivial}(b), from which we observe that the DAI criterion significantly reduces the  average number of queries over the trivial stopping.
We also observe that the simulated results closely match the proposed saddle-point approximation, confirming the accuracy and usefulness of the theoretical analysis.
\end{example}

\section{Algorithm Evaluation and Optimization}
In this section, we present performance analyses, comparisons of key metrics, and the mutual enhancement between GRAND and GCD. We also provide a brief review of the hardware implementations of GRAND in this section.

\subsection{Performance Gap to ML Decoding}
We have shown that with an unlimited number of guesses, both GRAND and GCD are MLD. Then an inevitable question arises: How does guessing decoding perform when the number of guesses is limited?
In this subsection, we analyze the performance gap between the guessing decoding algorithm and ML decoding when the number of guesses is limited to \(\ell_{\max}\).
Formally, the error event of the guessing decoding algorithm can be classified into two sub-events:
\begin{itemize}
  \item[\(E_0\):] The true TEP is included in the guessed patterns but is not the most likely one.  In this case ML decoding also fails. This sub-event cannot occur for the GRAND but does occur for the GCD.
  \item[\(E_1\):] The true TEP is not included in the first \(\ell_{\max}\) guesses, i.e., \(L(\bm\lambda)>\ell_{\max}\), where \(L(\bm\lambda)\) is the rank of the true TEP under the guessing order.
\end{itemize}
Hence, the overall FER is
\begin{equation}\label{eq:fer_gap}
  \mathrm{FER}
  = \mathbb{P}[E_0] + \mathbb{P}[E_1],
\end{equation}
and satisfies the bounds
\begin{equation}\label{eq:fer_bound}
\max\bigl(\varepsilon_{\mathrm{ML}},\!\mathbb{P}[L(\bm\lambda)\!>\!\ell_{\max}]\bigr)\!
\le \!
\mathrm{FER}
\le
\varepsilon_{\mathrm{ML}}\! +\! \mathbb{P}[L(\bm\lambda)\!>\!\ell_{\max}],
\end{equation}
where \(\varepsilon_{\mathrm{ML}}\) is the FER under ML decoding.  In particular,
\begin{equation}\label{eq:fer_upperbound_gap}
\mathrm{FER} - \varepsilon_{\mathrm{ML}}
\le
\mathbb{P}[L(\bm\lambda)>\ell_{\max}],
\end{equation}
so \(\mathbb{P}[L(\bm\lambda)>\ell_{\max}]\) quantifies the performance loss due to the limited search budget.
The tail probability \(\mathbb{P}[L(\bm\lambda)>\ell_{\max}]\) can be computed efficiently using the saddlepoint approximation described in Sec.~\ref{subseciton:GRAND-queries} and Sec.~\ref{subseciton:GCD-queries}.  For GRAND, we can obtain
$
\mathbb{P}[L_{\mathrm{GRAND}}(\bm\lambda)>\ell_{\max}]
$
via the approximation as \eqref{eq:GRAND-sd}.
For GCD, however, the relevant search set must reflect the rank of the true pattern among all valid candidates.  Hence, we redefine
\begin{equation}\label{Eq:searchset-gcd}
  \mathcal{L}_{\mathrm{C}}(\bm{\lambda})
  =
  \bigl\{\bm{f}\in\mathbb{F}_2^n : \Gamma_{\mathrm{S}}(\bm{f}_{\mathrm{I}})\le \Gamma_{\mathrm{S}}(\bm{e}_{\mathrm{I}}),\ \bm{f}_{\mathrm{P}}=\bm{0} |\bm{\lambda} \bigr\},
\end{equation}
where \(\bm{e}=(\bm{e}_{\mathrm{I}},\bm{e}_{\mathrm{P}})\) is the true TEP.  The probability that the true TEP falls outside the first \(\ell_{\max}\) patterns,
\(\mathbb{P}[L_{\mathrm{C}}(\bm\lambda)>\ell_{\max}]\), can then be approximated by the saddlepoint method.

\begin{figure}[tp]
    \centering
    \includegraphics[width=\figwidth]{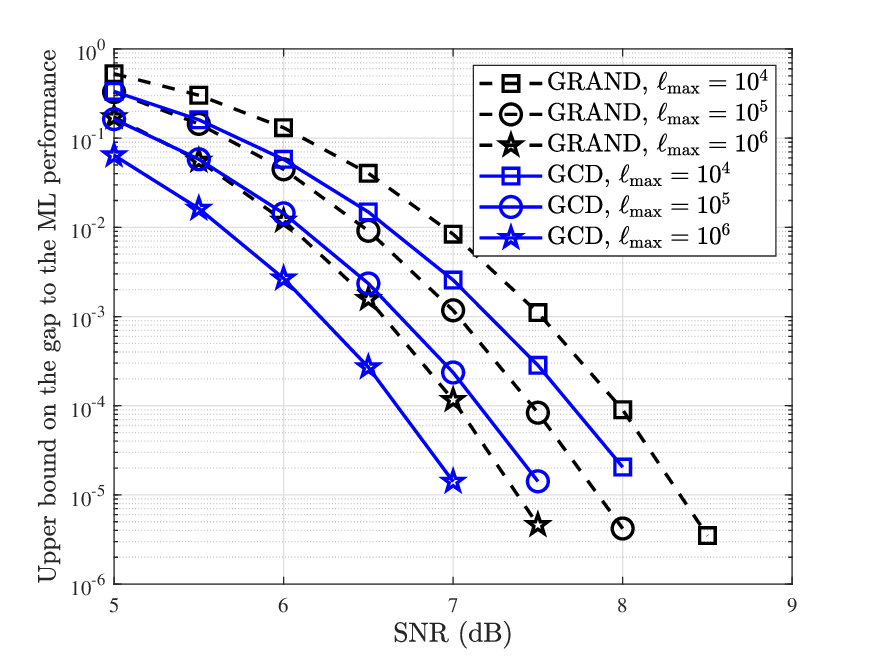}
    \caption{The upper bound on the performance gap to the ML decoding for the linear block code $\mathscr{C}[128,106]$ over BPSK-AWGN channels, where we consider GRAND~\cite{duffy2019capacity} and GCD~\cite{Zheng2024IT}.}
\label{fig:upper_bound_gap_ML_SNR_GCD}
\end{figure}

\begin{example}[Upper Bound on the Gap to ML Decoding]
Consider the linear block code \( \mathscr{C}[128,106] \) over BPSK-AWGN channels. The upper bounds on the gap to ML decoding for GRAND and GCD algorithms are shown in Fig.~\ref{fig:upper_bound_gap_ML_SNR_GCD}. From the figure, we observe that the upper bound on the gap to ML decoding decreases as either the SNR increases or the maximum number of queries \(\ell_{\max}\) increases. Moreover, the upper bound for GCD is slightly lower than that for GRAND across the tested settings.
\end{example}

\subsection{Key Metrics Comparison}\label{Subsection:KeyMetricComparison}

\subsubsection{Required Maximum Number of Queries}
Based on the tail probability \(\mathbb{P}[L(\bm\lambda)>\ell_{\max}]\), one can choose \(\ell_{\max}\) to achieve near-ML performance with controlled worst-case complexity.  A straightforward rule of thumb is that if the gap to ML decoding is sufficiently small (e.g., an order of magnitude smaller compared to the case without a search limit), the performance loss (in terms of FER) caused by the limited  searches can be neglected.
Here, for simplicity, the finite-length bound can be used as a substitute for ML decoding performance. Based on this, we can choose a minimum required $\ell_{\max}$, denoted as $\widetilde{\ell}_{\max}$, such that
\begin{equation}\label{Eq:determin-ell-max_FPT-BM}
\mathbb{P}[L(\bm \lambda) > \widetilde{\ell}_{\max} ] \le  \alpha\varepsilon', \quad \varepsilon' = {\varepsilon}_{\textrm{ML}} \mbox{ or } \varepsilon_{\mathrm{RCU}},
\end{equation}
where $\alpha \in (0,1]$ and $\varepsilon_{\mathrm{RCU}}$ is the RCU bound for
codes of length $n$ and dimension $k$.
The resulting \(\widetilde\ell_{\max}\) for GRAND and GCD is illustrated in the following example.

\begin{example}[Minimum Required $\widetilde{\ell}_{\max}$ versus Code Rate]\label{ex:rcu_ell_max}
Consider binary codes transmitted over BPSK-AWGN channels.  Using the RCU bound, we first determine, for each code rate \(r\), the SNR needed to achieve \(\varepsilon_{\mathrm{RCU}}=10^{-5}\).  We then apply the saddlepoint approximation together with the criterion in \eqref{Eq:determin-ell-max_FPT-BM} (with \(\alpha=1\)) to compute the minimum required search budget \(\widetilde{\ell}_{\max}\) for a given code rate and the determined SNR targeting \(\varepsilon_{\mathrm{RCU}}\).  Fig.~\ref{fig:ellmax_and_queries}(a) plots \(\widetilde{\ell}_{\max}\) for different code rates.  From this figure we observe that $\widetilde{\ell}_{\max}$ of GRAND grows roughly exponentially as the code rate \(r\) decreases.
We also observe that GCD requires a smaller \(\widetilde{\ell}_{\max}\) than GRAND, especially in the low-rate regime.
\end{example}


\subsubsection{Average Number of Queries}
The saddlepoint approximation provides a rapid and accurate estimate of the average list size (i.e., number of queries) for both GRAND and GCD. We compare these results over two channels: the binary symmetric channel (BSC) and the BPSK-AWGN channel.

\begin{example}[BSC Channel]
\label{ex:bsc-num}
Consider a binary linear code \(\mathscr{C}[128,105]\) over a BSC with crossover probability \(p\).  Here the soft weight reduces to Hamming weight, and the analytic upper bounds on the number of guesses are
\begin{equation}\label{eq:bsc-analytic}
\overline{L}_{\mathrm{N}}
=\sum_{w=0}^{\Gamma_{\mathrm{H}}(\bm{e})}\binom{n}{w},
\quad
\overline{L}_{\mathrm{C}}
=\sum_{w=0}^{\Gamma_{\mathrm{H}}(\bm{e})}\binom{k}{w}.
\end{equation}
Table~\ref{tab:BSC-list-size} compares these exact bounds to their saddlepoint approximations.  Even for small weights, the relative error stays below \(8\%\), and it rapidly decreases as \(\Gamma_{\mathrm{H}}(\bm{e})\) grows, confirming the method’s accuracy.
\end{example}

\begin{table}[!t]
  \centering
  \caption{Exact vs.\ approximate list sizes for \(\mathscr{C}[128,105]\) over BSC.}
  \label{tab:BSC-list-size}
  \resizebox{\columnwidth}{!}{%
  \begin{tabular}{c|ccc|ccc}
    \toprule
    \multirow{2}{*}{\(\Gamma_{\mathrm{H}}(\bm{e})\)} &
    \multicolumn{3}{c|}{GRAND (\(\overline{L}_{\mathrm{N}}\))} &
    \multicolumn{3}{c}{GCD (\(\overline{L}_{\mathrm{C}}\))} \\
    \cmidrule{2-7}
    & Exact & Approx. & Error & Exact & Approx. & Error \\
    \midrule
    1  & \(1.29\!\times\!10^2\)  & \(1.39\!\times\!10^2\)  & \(+7.8\%\)  & \(1.06\!\times\!10^2\)  & \(1.14\!\times\!10^2\)  & \(+7.6\%\) \\
    2  & \(8.26\!\times\!10^3\)  & \(8.58\!\times\!10^3\)  & \(+3.9\%\)  & \(5.57\!\times\!10^3\)  & \(5.78\!\times\!10^3\)  & \(+3.8\%\) \\
    4  & \(1.10\!\times\!10^7\)  & \(1.12\!\times\!10^7\)  & \(+1.7\%\)  & \(4.97\!\times\!10^6\)  & \(5.06\!\times\!10^6\)  & \(+1.6\%\) \\
    8  & \(1.53\!\times\!10^{12}\) & \(1.54\!\times\!10^{12}\) & \(+0.7\%\)  & \(3.03\!\times\!10^{11}\) & \(3.05\!\times\!10^{11}\) & \(+0.6\%\) \\
    \bottomrule
  \end{tabular}
  }
\end{table}

For BPSK-AWGN channels, closed-form expressions for queries on average are unavailable, but the saddle-point approximation still applies.  Moreover, for a random code of rate \(r=k/n\), the RCU bound can be written as~\cite{Font2018Saddlepoint}
\begin{equation}\label{eq:rcu}
\varepsilon_{\mathrm{RCU}}
\!=\!\mathbb{E}\Bigl[\min\bigl\{1,\,(2^k\!-\!1)\,\mathbb{P}\bigl[\Gamma_{\mathrm{S}}(\bm{f})\!\le\!\Gamma_{\mathrm{S}}(\bm{e})\!\mid\!\bm{\lambda}\bigr]\bigr\}\Bigr],
\end{equation}
where \(\mathbb{P}[\Gamma_{\mathrm{S}}(\bm{f})\le\Gamma_{\mathrm{S}}(\bm{e})\mid\bm{\lambda}]\) is the probability that a random pattern \(\bm{f}\) has lower soft weight than the true error pattern \(\bm{e}\).  Since
\begin{equation}
\overline{L}_{\mathrm{N}}
=2^n\,\mathbb{P}\bigl[\Gamma_{\mathrm{S}}(\bm{f})<\Gamma_{\mathrm{S}}(\bm{e})\bigr],
\end{equation}
a simple rearrangement yields the lower bound
\begin{equation}\label{equ:low_bound_ln}
\overline{L}_{\mathrm{N}}
\ge
2^{\,n-k}\,\varepsilon_{\mathrm{RCU}}.
\end{equation}
This bound directly links the average number of GRAND's search steps to the RCU bound, thereby quantifying the fundamental trade-off between complexity and reliability.

\begin{example}[Average Queries versus Code Rate]\label{ex:rcu}

Consider binary codes over BPSK-AWGN channels.
Using the RCU bound, we can first determine the SNR required for a random code to achieve \(\varepsilon_{\mathrm{RCU}}=10^{-5}\).  Then, for each code rate \(r\) and the corresponding determined SNR, we estimate the average number of guesses via the saddlepoint method.  Fig.~\ref{fig:ellmax_and_queries}(b) shows both the saddlepoint estimates of \(\overline{L}_{\mathrm{N}}\) and the lower bound \(2^{n-k}\varepsilon_{\mathrm{RCU}}\).  We observe that: 1) \(\overline{L}_{\mathrm{N}}\) grows exponentially as code rate \(r\) decreases; 2) For \(n=128\), codes of rate below \(3/4\) require more than \(10^5\) guesses, making GRAND impractical; 3) GCD always requires fewer guesses than GRAND, especially at low rates.
\end{example}

\begin{figure}[tp]
  \centering
  \subfloat[Minimum required $\widetilde{\ell}_{\max}$]{%
    \label{fig:maximum_queries_GRAND_GCD}
    \resizebox{0.45\columnwidth}{!}{
%
%
\definecolor{mycolor1}{rgb}{0.46600,0.67400,0.18800}%
\definecolor{mycolor2}{rgb}{0.00000,0.44700,0.74100}%
\definecolor{mycolor3}{rgb}{0.49400,0.18400,0.55600}%
\definecolor{mycolor4}{rgb}{0.85000,0.32500,0.09800}%
\begin{tikzpicture}[%
thick,scale=1.1, every node/.style={scale=1.1}
]

\begin{axis}[%
width=10.4cm,
height=7.6781158cm,
at={(0cm,0cm)},
scale only axis,
xmin=0,
xmax=1,
xlabel style={font=\color{white!15!black}},
xlabel={Code rate},
ymode=log,
ymin=1,
ymax=1e+30,
ytick={               1,           1e5,      1e10, 1e15, 1e+20, 1e+25, 1e+30},
xtick={0, 0.125, 0.25, 0.375, 0.5, 0.625, 0.75, 0.875, 1},
xticklabel style={
  /pgf/number format/fixed,
  /pgf/number format/precision=3
},
yminorticks=true,
ylabel style={font=\color{white!15!black}},
ylabel={ \(\widetilde\ell_{\max}\)},
axis background/.style={fill=white},
xmajorgrids,
ymajorgrids,
yminorgrids,
legend style={legend cell align=left, align=left, draw=white!15!black},
legend columns=1
]

\addplot [color=mycolor2, densely dotted, line width=1.4pt, mark size=3.5pt, mark=+, mark options={solid, mycolor2}]
  table[row sep=crcr]{%
0.125000 6.133e32\\
0.250000 1.287e28\\
0.375000 2.098e23\\
0.500000 3.120e18\\
0.625000 4.4e13\\
0.750000 4.6e8\\
0.875000 613\\
};
\addlegendentry{GRAND, $n = 128$}

\addplot [color=mycolor4, dash=on 8pt off 5pt phase 0pt, line width=1.4pt, mark size=3.1pt, mark=o, mark options={solid, mycolor4}]
  table[row sep=crcr]{%
0.125000 32691 \\
0.250000 2.660e8 \\
0.375000 2.566e10 \\
0.500000 6.069e10 \\
0.625000 4.849e9 \\
0.750000 1.371e7 \\
0.875000 380 \\
};
\addlegendentry{GCD, $n = 128$}

\end{axis}
\end{tikzpicture}%
}%
  }
  \hfill
  \subfloat[Average number of queries]{%
    \label{fig:queries_GRAND_GCD}
    \resizebox{0.45\columnwidth}{!}{
%
%
\definecolor{mycolor1}{rgb}{0.46600,0.67400,0.18800}%
\definecolor{mycolor2}{rgb}{0.00000,0.44700,0.74100}%
\definecolor{mycolor3}{rgb}{0.49400,0.18400,0.55600}%
\definecolor{mycolor4}{rgb}{0.85000,0.32500,0.09800}%
\begin{tikzpicture}[%
thick,scale=1.1, every node/.style={scale=1.1}
]

\begin{axis}[%
width=10.4cm,
height=7.6781158cm,
at={(0cm,0cm)},
scale only axis,
xmin=0,
xmax=1,
xlabel style={font=\color{white!15!black}},
xlabel={Code rate},
ymode=log,
ymin=1,
ymax=1e+30,
ytick={               1,           1e5,      1e10, 1e15, 1e+20, 1e+25, 1e+30},
xtick={0, 0.125, 0.25, 0.375, 0.5, 0.625, 0.75, 0.875, 1},
xticklabel style={
  /pgf/number format/fixed,
  /pgf/number format/precision=3
},
yminorticks=true,
ylabel style={font=\color{white!15!black}},
ylabel={Average number of queries},
axis background/.style={fill=white},
xmajorgrids,
ymajorgrids,
yminorgrids,
legend style={legend cell align=left, align=left, draw=white!15!black},
legend columns=1
]

\addplot [color=mycolor2, line width=1.4pt]
  table[row sep=crcr]{%
0.078125 3.32306998946229e+30\\
0.125	5.19229685853483e+28\\
0.25	7.92281625142643e+23\\
0.375	1.20892581961463e+19\\
0.5	184467440737096\\
0.625	2814749767.10656\\
0.75	42949.67296\\
0.875	0.65536\\
};
\addlegendentry{(\ref{equ:low_bound_ln}), $n = 128$}

\addplot [color=mycolor2, densely dotted, line width=1.4pt, mark size=3.5pt, mark=+, mark options={solid, mycolor2}]
  table[row sep=crcr]{%
0.078125 4.52306998946229e+30\\
0.125	7.4e+28\\
0.25	1.47e+24\\
0.375	2.63e+19\\
0.5	398000000000000\\
0.625	5310000000\\
0.75	66400\\
0.875	1.29\\
};
\addlegendentry{GRAND, $n = 128$}

\addplot [color=mycolor4, dash=on 8pt off 5pt phase 0pt, line width=1.4pt, mark size=3.1pt, mark=o, mark options={solid, mycolor4}]
  table[row sep=crcr]{%
0.0625	1.99\\
0.125	19200\\
0.1875	6110000\\
0.25	184000000\\
0.3125	1340000000\\
0.375	3000000000\\
0.4375	2510000000\\
0.5	835000000\\
0.5625	123000000\\
0.625	8890000\\
0.6875	324000\\
0.75	7120\\
0.8125	71.8\\
0.875	1.24\\
0.9375	0\\
};
\addlegendentry{GCD, $n = 128$}

\end{axis}
\end{tikzpicture}%
}%
  }
  \caption{%
   (a) Minimum required $\widetilde{\ell}_{\max}$ vs.\ code rate;
  (b) average number of queries vs.\ code rate.
  Results for $n=128$ over BPSK-AWGN, targeting $\varepsilon_{\mathrm{RCU}}=10^{-5}$.}
  \label{fig:ellmax_and_queries}
\end{figure}


\subsubsection{Average Complexity Analysis}
While \(\overline{L}_{\mathrm{N}}\) and \(\overline{L}_{\mathrm{C}}\) quantify the guesses on average, actual complexity also depends on per-guess cost. Here we quantify the computational cost per guess and compare overall complexity.

\paragraph{GRAND}
Each query in GRAND checks parity-equations.  Assuming that each invalid pattern fails the parity-check with probability $1/2$, the expected number of row checks per guess is
$
 \sum_{i=1}^{m} i \left(\tfrac12\right)^i \lesssim 2,
$
with \(m=n-k\).  A single row-check requires \(2n-1\) operations (one length-\(n\) dot product).  Hence, the average number of operations for GRAND is
\begin{equation}\label{Eq:GRAND-ops}
\mathrm{Ops}_{\mathrm{GRAND}}
=
\bigl(\overline{L}_{\mathrm{N}}\!-\!1\bigr)\!\times\! 2\!\times\!(2n\!-\!1)
\!\!+\!\!(n\!-\!k)\,(2n\!-\!1).
\end{equation}

\paragraph{GCD}
Each guess in GCD re-encodes the parity part at cost \((n-k)\) dot-products of length \(k\), i.e., \((n-k)(2k-1)\) operations.  Thus, the average number of operations for GCD is
\begin{equation}\label{Eq:GCD-ops}
\mathrm{Ops}_{\mathrm{GCD}}
=
\overline{L}_{\mathrm{C}}\times (n-k)\,(2k-1).
\end{equation}

\begin{figure}[tp]
  \centering
  \resizebox{0.35\textwidth}{!}{
%
%
\definecolor{mycolor1}{rgb}{0.46600,0.67400,0.18800}%
\definecolor{mycolor2}{rgb}{0.00000,0.44700,0.74100}%
\definecolor{mycolor3}{rgb}{0.49400,0.18400,0.55600}%
\definecolor{mycolor4}{rgb}{0.85000,0.32500,0.09800}%
\begin{tikzpicture}[%
thick,scale=1.1, every node/.style={scale=1.1}
]

\begin{axis}[%
width=10.4cm,
height=7.6781158cm,
at={(0cm,0cm)},
scale only axis,
xmin=0,
xmax=1,
xlabel style={font=\color{white!15!black}},
xlabel={Code rate},
ymode=log,
ymin=1,
ymax=1e+30,
ytick={               1,           1e5,      1e10, 1e15, 1e+20, 1e+25, 1e+30},
xtick={0, 0.125, 0.25, 0.375, 0.5, 0.625, 0.75, 0.875, 1},
xticklabel style={
  /pgf/number format/fixed,
  /pgf/number format/precision=3
},
yminorticks=true,
ylabel style={font=\color{white!15!black}},
ylabel={Average number of operations},
axis background/.style={fill=white},
xmajorgrids,
ymajorgrids,
yminorgrids,
legend style={legend cell align=left, align=left, draw=white!15!black},
legend columns=1
]
\addplot [color=mycolor2, line width=1.4pt]
  table[row sep=crcr]{%
0.078125	1.6947657E+33 \\
0.125	2.6480714E+31 \\
0.25	4.0406363E+26 \\
0.375	6.1655217E+21 \\
0.5	9.4078395E+16 \\
0.625	1.4355224E+12 \\
0.75	2.1911983E+07 \\
0.875	3.9042336E+03 \\
};
\addlegendentry{(\ref{equ:low_bound_ln}) and \eqref{Eq:GRAND-ops}, $n = 128$}

\addplot [color=mycolor2, densely dotted, line width=1.4pt, mark size=3.5pt, mark=+, mark options={solid, mycolor2}]
  table[row sep=crcr]{%
0.078125	2.3067657E+33 \\
0.125	3.7740000E+31 \\
0.25	7.4970000E+26 \\
0.375	1.3413000E+22 \\
0.5	2.0298000E+17 \\
0.625	2.7081000E+12 \\
0.75	3.3871650E+07 \\
0.875	4.2279000E+03 \\
};
\addlegendentry{GRAND by \eqref{Eq:GRAND-ops}, $n = 128$}

\addplot [color=mycolor4, dash=on 8pt off 5pt phase 0pt, line width=1.4pt, mark size=3.1pt, mark=o, mark options={solid, mycolor4}]
  table[row sep=crcr]{%
0.0625	1.2141547E+05 \\
0.125	1.0569644E+13 \\
0.1875	9.9392983E+17 \\
0.25	8.3204506E+20 \\
0.3125	4.0451277E+22 \\
0.375	1.8432000E+23 \\
0.4375	1.1612344E+23 \\
0.5	1.1423334E+22 \\
0.5625	2.1688934E+20 \\
0.625	9.7114644E+17 \\
0.6875	1.0749542E+15 \\
0.75	4.1528830E+11 \\
0.8125	3.1672071E+07 \\
0.875	6.2781696E+03 \\
0.9375	0.0000000E+00 \\
};
\addlegendentry{GCD by \eqref{Eq:GCD-ops}, $n = 128$}

\end{axis}
\end{tikzpicture}%
}
  \caption{Average operations vs.\ code rate over BPSK-AWGN with $n = 128$, targeting \(\varepsilon_{\mathrm{RCU}}=10^{-5}\).}
  \label{fig:operations_GRAND_GCD}
\end{figure}

\begin{example}[Average Operations versus Code Rate]
Consider binary codes of length \(n\) and varying rate \(r=k/n\) over BPSK-AWGN channels.  Using the saddlepoint estimates of \(\overline{L}_{\mathrm{N}}\) and \(\overline{L}_{\mathrm{C}}\) from Secs.~\ref{subseciton:GRAND-queries} and \ref{subseciton:GCD-queries} and the lower bound of \(\overline{L}_{\mathrm{N}}\) as~\eqref{equ:low_bound_ln}, we compute \(\mathrm{Ops}_{\mathrm{GRAND}}\) and \(\mathrm{Ops}_{\mathrm{GCD}}\) via \eqref{Eq:GRAND-ops} and \eqref{Eq:GCD-ops}, as shown in Fig.~\ref{fig:operations_GRAND_GCD}. From this figure we observe that: 1) At high code rates, both GRAND and GCD exhibit comparable average computational effort, as both algorithms require only very small numbers of average queries; 2) At moderate to high rates, GRAND is more efficient, since its early-exit parity checks sharply reduce the cost per guess; 3) At low rates, GCD prevails, as its average number of guesses \(\overline{L}_{\mathrm{C}}\) becomes much smaller than \(\overline{L}_{\mathrm{N}}\).
\end{example}

\subsection{Mutual Promotion of GRAND and GCD}

GRAND and GCD share the same ``guess-and-check'' philosophy, and innovations for one often translate directly to the other for improving the performance or reducing the complexity. We highlight three key examples.

\subsubsection{ORB-Based Integerization}
ORBGRAND~\cite{duffy2022ordered} replaces real-valued LLRs with their rank indices \(\{\Lambda_i\}\) and enumerates patterns by the logistic weight
$
\Gamma_{\mathrm{L}}(\bm e)
=\sum_{i=1}^n \Lambda_i\,e_i,
$
achieving near-soft-weight performance using only integer operations. This idea extends naturally to GCD. Specifically, assign each information-bit position \(i\) an integer reliability \(\Lambda_i\) and define
$
\Gamma_{\mathrm{L}}(\bm{e}_{\mathrm{I}})
=\sum_{i=1}^k \Lambda_i\,e_i.
$
Then enumerate \(\bm{e}_{\mathrm{I}}\) in non-decreasing order of \(\Gamma_{\mathrm{L}}(\bm{e}_{\mathrm{I}})\) and re-encode to obtain the corresponding parity bits~\cite{Wang2024ordered}.

\subsubsection{Local Constraint}
LC-OSD~\cite{Wang2022LCOSD,liang2023randomarXiv} uses partial GE to impose \(\delta\) additional syndrome checks. By permuting the parity-check matrix \(\mathbf{H}\) into the form
\begin{equation}
  \label{equ:gauss-elim}
  \widetilde{\mathbf{H}}=
  \begin{bNiceArray}{cw{c}{2cm}|cw{c}{1cm}}[margin, first-row, last-col]
    \Block{1-2}{_{n-k-\delta}} & & \Block{1-2}{_{k+\delta}} \\
    \Block{3-2}{\mathbf{I}} & & \Block{3-2}{\mathbf{P}_1} & & \Block{3-1}{^{\rotate n-k-\delta}} \\
    & & & \\
    & & & \\
    \hline
    \Block{1-2}{\mathbf{O}} & & \Block{1-2}{\mathbf{P}_2} & & \Block{1-1}{^{\rotate \delta}} \\
  \end{bNiceArray}
\end{equation}
a permuted error pattern \(\widetilde{\bm e} = (\widetilde{\bm e}_L, \widetilde{\bm e}_R)\), with \(\widetilde{\bm e}_L \in \mathbb{F}_2^{n-k-\delta}\) and \(\widetilde{\bm e}_R \in \mathbb{F}_2^{k+\delta}\), must satisfy
\begin{align}
  \widetilde{\bm e}_L + \widetilde{\bm e}_R\,\mathbf{P}_1^\mathsf{T} &= \widetilde{\bm s}_1
    \triangleq \widetilde{\bm z}_L + \widetilde{\bm z}_R\,\mathbf{P}_1^\mathsf{T}, \label{equ:el-parity}\\
  \widetilde{\bm e}_R\,\mathbf{P}_2^\mathsf{T} &= \widetilde{\bm s}_2
    \triangleq \widetilde{\bm z}_R\,\mathbf{P}_2^\mathsf{T}. \label{equ:er-parity}
\end{align}
Thus, one only needs to enumerate \(\widetilde{\bm e}_R\in\mathbb{F}_2^{k+\delta}\) in increasing partial soft weight satisfying~\eqref{equ:er-parity}, and then compute \(\widetilde{\bm e}_L\) using~\eqref{equ:el-parity}.

In LC-GCD~\cite{Zheng2024ITW}, selecting \(\delta\) rows imposes these extra checks during the information-bit guessing phase, effectively pruning invalid candidates early and reducing complexity. Particularly, SPC-GCD embeds a single-parity-check constraint to eliminate many re-encodings~\cite{griffin2025using}. The same principle applies to GRAND: PC-GRAND and constrained GRAND incorporate a small number of syndrome checks during guessing to accelerate decoding~\cite{wang2024partially,rowshan2022constrained}.

\subsubsection{Soft-Output Decoding}

Soft-output (SO) decoders produce reliability metrics—block posteriors and bit LLRs—vital for hybrid-ARQ and iterative decoding.  Unlike traditional methods that require list decoding or multiple passes~\cite{forney1968exponential}, GRAND and GCD yield SO “for free” via their single-pass guessing~\cite{yuan2025soft}.  Moreover, their soft-input soft-output~(SISO) extensions enable efficient decoding of tensor-product codes up to dimensions $[n^2,k^2]$.

\paragraph{GRAND SO}
Let $\bm y\in\mathbb R^n$ and $\{\bm e^{(i)}\}$ be noise guesses in descending likelihood.  If the first $L$ valid codewords occur at indices $q_1<\cdots<q_L$, then
\[
\bm c^{(j)}=\bm y\oplus\bm e^{(q_j)},\quad j=1,\dots,L,
\]
and one approximates
\begin{equation*}\label{eq:SOGRAND}
\mathbb{P}(\bm c^{(j)}|\bm y)
\!\approx\!
\frac{P(\bm e^{(q_j)}|\bm y)}
{\sum_{i=1}^L P(\bm e^{(q_i)}|\bm y)
\!+\!\Bigl(1\!\!-\!\!\sum_{i=1}^{q_L}P(\bm e^{(i)}|\bm y)\Bigr)\frac{2^k-1}{2^n-1}}.
\end{equation*}
Bit LLRs follow by summing these block posteriors over codewords with the bit equal to $0$ or $1$.

\paragraph{GCD SO}
GCD can be viewed as guessing patterns \(\bm e_{\mathrm I}\in\mathbb{F}_2^k\) on the information bits and then re-encoding to full codewords.
Let the first \(L\) successful GCD re-encodings occur at indices
\(\{q_1,\dots,q_L\}\) with $q_i = i$, and define
\[
P_{\mathrm I}(i)
=\sum_{\bm a\in\mathbb{F}_2^{n-k}}
P\bigl((\bm e_{\mathrm I}^{(i)},\bm a)\mid\bm y\bigr).
\]
Then
\begin{equation}\label{eq:SOGCD_block}
\mathbb{P}(\bm c^{(j)}|\bm y)
\approx
\frac{P_{\mathrm I}(j)}
{\sum_{i=1}^L P_{\mathrm I}(i)
+\Bigl(1-\sum_{i=1}^L P_{\mathrm I}(i)\Bigr)\frac{2^k-1}{2^n-1}}.
\end{equation}
Bit LLRs combine these posteriors and the “not-in-list” residual.
Under proper scoring rules, these SO estimates closely approach true ML posteriors~\cite{griffin2025using}.

\subsection{Towards Hardware Implementation}

GRAND has been demonstrated in high-throughput, low-energy silicon implementations, making it well suited for ultra-reliable, low-latency communications. The first synthesizable GRAND decoder~\cite{abbas2020high} used a dial structure to reduce both query count and latency.  A universal hard-detection GRAND core supporting multi-code, multi-rate decoding (up to 128 bits, rate $\ge0.66$) appeared in~\cite{Riaz2022}, and a VLSI architecture optimized for channels with memory was introduced in~\cite{Abbas2021}.  For integer-based decoding, ORBGRAND architectures were proposed in~\cite{abbas2022high}, followed by a fixed-latency, LUT-based design in~\cite{condo2022fixed}.  Most recently, a 40 nm CMOS implementation achieved 6.5 Gb/s at 0.76 pJ/bit and 4.9 mW~\cite{riaz2024sub}, and a 65 nm CMOS version reached a worst-case throughput of 8.48 Gb/s~\cite{Blanc2024}.  A 28 nm ASIC for hard-decision GRAND also demonstrated area- and power-efficient decoding for systematic codes~\cite{Chu2023}.

Since GCD shares GRAND's core ``guess-and-check'' mechanism—differing mainly by flipping information bits and re-encoding—the same hardware techniques apply\,:  ORBGRAND’s integer ranking, LUT-based schedulers, and early-stopping logic can be directly adapted to support the information-bit search in GCD. Moreover, the pipelined datapaths used for parity checks in GRAND can be reused for systematic re-encoding in GCD. We therefore expect that optimized GCD cores can achieve similar silicon efficiency, enabling high-speed, low-power decoding  for low-rate codes.

\section{Conclusion and Future Directions}

In this paper, we have provided a unified and comprehensive study of guessing decoding for short blocklength codes, focusing on two complementary families: GRAND and GCD.  Our main contributions are as follows:
\begin{itemize}
  \item We detailed the algorithmic implementations of GRAND and GCD, including a variety of pattern-generation orders (soft-weight, ORB, etc.) and early-stopping criteria.
  \item We proved that, under appropriate ordering and stopping rules, both GRAND and GCD achieve maximum-likelihood decoding.
  \item We derived saddle-point approximations for the average number of queries required by each decoder, and validated these approximations against simulation results.
  \item We quantified the performance gap to ML decoding under a limited search budget, compared key complexity metrics (worst-case and average query counts, per-query operations, and hardware implications), delineated the rate regimes in which each method excels.
  \item We surveyed recent VLSI developments for GRAND and discussed the prospects for hardware implementation of GCD.
\end{itemize}

Looking forward, we identify several promising avenues for further research in guessing decoding:

\begin{enumerate}
  \item \textbf{Theoretical Analysis:}  While query-count distribution under soft-weight ordering of guessing decoding is now well understood via saddle-point techniques, the analysis for guessing decoding under ORB ordering or other integer-rank approximations remains an open problem.  Extending large-deviation or saddle-point methods to these practical orders will deepen complexity-performance trade-off of guessing decoding.

  \item \textbf{Guessing Paradigms:} Current decoders generate patterns on the fly using FPT (soft-weight) or ORB (logistic-weight) criteria.  Future work could investigate hybrid offline-online schemes in which a small set of highly probable patterns is precomputed (e.g.\ code- or channel-dependent lookup tables) and then complemented by real-time pattern generation and early stopping.  Such hybrid paradigms may yield further complexity reductions without sacrificing performance.

  \item \textbf{Broader Applications:}  Beyond moderate-redundancy or moderate-dimension short codes, guessing decoders could be adapted to more complex settings, including:
    \begin{itemize}
      \item Decoding of long codes by applying guessing decoding to short component codes or subcodes—e.g., product codes and polar codes—or by integrating it into iterative LDPC decoding. In addition to wireless communications, similar applications in storage systems could also be explored, for example for decoding short embedded blocks within larger code structures.
      \item Multiuser and multi-access scenarios, where guessing decoding can be combined with joint-detection or interference-cancellation techniques.
      \item Application-specific optimizations for ultra-low-power or latency-critical links in 6G and beyond.
    \end{itemize}

  \item \textbf{Hardware Architectures for GRAND and GCD:}
    GRAND's VLSI implementations have matured rapidly, culminating in sub-pJ/bit 40nm ORBGRAND ASICs.  Future work can explore novel microarchitectures—such as deeply pipelined syndrome-check engines, on-chip ORB schedulers, and memory-efficient lookup tables—to push throughput and energy efficiency even further.  By contrast, GCD’s hardware support remains nascent; developing compact, fixed-latency cores that exploit its re-encoding structure and partial-constraint pruning will be key to crafting truly universal, high-speed decoding engines that seamlessly switch between GRAND and GCD modes.

\end{enumerate}



\bibliographystyle{IEEEtran}
\bibliography{bibliofile}


%
%
%
%
%
%
%

\end{document}